\documentclass[a4paper, 12pt]{article}

\usepackage{fullpage}
\usepackage{authblk}
\usepackage{amsmath}
\usepackage{xfrac}
\usepackage{url}
\usepackage{caption}
\usepackage{csquotes}

\MakeOuterQuote{"}

\title{Exceedance Probabilities \\ for the Dirichlet Distribution}
\author[1,3]{Joram Soch}
\author[1,2]{Carsten Allefeld}
\affil[1]{Bernstein Center for Computational Neuroscience, Berlin, Germany}
\affil[2]{Berlin Center for Advanced Neuroimaging, Berlin, Germany}
\affil[3]{Department of Psychology, Humboldt-Universität zu Berlin, Germany}
\date{}

\begin{document}

\setcounter{page}{0}
\vspace*{1em}
\begin{center}
	\LARGE
		Exceedance Probabilities \\ for the Dirichlet Distribution \\ \vspace{1em}
	\large
		Joram Soch\textsuperscript{1,3,\textbullet} \& Carsten Allefeld\textsuperscript{1,2}
\end{center}
\begin{flushleft}
	\normalsize
		\hspace{2.5em}
		\textsuperscript{1} Bernstein Center for Computational Neuroscience, Berlin, Germany \\
		\hspace{2.5em}
		\textsuperscript{2} Berlin Center for Advanced Neuroimaging, Berlin, Germany \\
		\hspace{2.5em}
		\textsuperscript{3} Department of Psychology, Humboldt-Universit\"{a}t zu Berlin, Germany \\ \vspace{1em}
		\hspace{2.5em}
		\textbullet \ Corresponding author: \url{joram.soch@bccn-berlin.de}.		
\end{flushleft}
\vspace*{1em}

\begin{abstract}
\noindent
We derive an efficient method to calculate exceedance probabilities (EP) for the Dirichlet distribution when the number of event types is larger than two. Also, we present an intuitive application of Dirichlet EPs and compare our method to a sampling approach which is the current practice in neuroimaging model selection.
\end{abstract}

\vspace*{1em}
\tableofcontents

\pagebreak
\section{Introduction}

Let $r = [r_1, \ldots, r_k]$ be a $1 \times k$ random vector. Then, $r$ is said to follow a \textit{Dirichlet distribution}, if its probability density function is given by (Gelman et al., 2013, p.~579)

\begin{equation} \label{eq:Dir-pdf}
p(r) = \mathrm{Dir}(r; \alpha) = \frac{\Gamma(\sum_{j=1}^k \alpha_i)}{\prod_{j=1}^k \Gamma(\alpha_j)} \, \prod_{j=1}^k r_j^{\alpha_j-1}
\end{equation}

where $\alpha_1, \ldots, \alpha_k$ are called concentration parameters and $r$ underlies the constraint that

\begin{equation} \label{eq:Dir-const}
0 \leq r_j \leq 1 \quad \text{for} \quad j = 1, \ldots, k \quad \text{and} \quad \sum_{j=1}^k r_j = 1 \; .
\end{equation}

Let $E_1, \ldots, E_k$ be some mutually exclusive and collectively exhaustive event types with the unknown occurence frequencies $r_1, \ldots, r_k$. Then, $\mathrm{Dir}(r; \alpha)$ as given by equation (\ref{eq:Dir-pdf}) describes the probability of each frequency combination satisfying the constraint (\ref{eq:Dir-const}). Since $\alpha = [1, ..., 1]$ invokes a flat distribution over $r$, $\alpha_j - 1$ can be seen as the number of pseudo-observations of event type $E_j$. This is also reflected in the mode of the Dirichlet which is (Gelman et al., 2013, p.~579)

\begin{equation} \label{eq:Dir-mode}
\mathrm{mode}(r_j) = \frac{\alpha_j - 1}{\alpha_s - k} \quad \text{with} \quad \alpha_s = \sum_{j=1}^k \alpha_j \; .
\end{equation}

The Dirichlet distribution is often used as the conjugate prior for a multinomial likelihood (Gelman et al., 2013, pp.~578, 584) which gives rise to a Multinomial-Dirichlet model (see Section \ref{sec:App_PEF}), a generalization of the Binomial-Beta model constituted by a binomial distribution and a beta distribution.

\vspace{0.5em}

Let $X$ and $Y$ be two continuous random variables with the joint probability density function $p(x,y)$. Then, the probability

\begin{equation} \label{eq:EP-def}
p(x>y) = \iint_{x>y} p(x,y) \, \mathrm{d}x \, \mathrm{d}y
\end{equation}

is referred to as an \textit{exceedance probability} (EP) (Stephan et al., 2009, p.~1008), more specifically the probability that $X$ exceeds $Y$. EPs are a useful tool for posterior parameter inference in Bayesian statistics, since a lot of interesting questions can be understood as questions for exceedance events. For example, if $X$ and $Y$ are independent and $X \sim \mathrm{N}(\mu_x,\sigma_x)$ and $Y \sim \mathrm{N}(\mu_y,\sigma_y)$, then

\begin{equation} \label{eq:N-N-EP}
\begin{split}
p(x>y) &= \int_{-\infty}^{+\infty} \int_{y}^{\infty} \mathrm{N}(x; \mu_x,\sigma_x) \, \mathrm{N}(y; \mu_y,\sigma_y) \, \mathrm{d}x \, \mathrm{d}y \\
&= 1 - \left\langle \Phi_{\mu_x,\sigma_x}(y) \right\rangle
\end{split}
\end{equation}

where $\Phi_{\mu,\sigma}$ is the cumulative distribution function of the normal distribution $\mathrm{N}(\mu,\sigma)$. EPs for the Dirichlet distribution become especially important in group-level Bayesian model selection (see Section \ref{sec:App_NMS}).

\pagebreak
\section{Theory}

\subsection{Dirichlet exceedance probabilities} \label{sec:Dir_EPs}

For a Dirichlet distribution $\mathrm{Dir}(r; \alpha)$, as defined by equation (\ref{eq:Dir-pdf}), the exceedance probability $\varphi_j$ is understood as the probability that $r_j$ is larger than all other elements of $r$ (Stephan et al., 2009, p.~1008):

\begin{equation} \label{eq:Dir-EP-def}
\varphi_j = p \left( \forall i \in \left\lbrace 1,\ldots,k | j \neq k \right\rbrace: \, r_j > r_i |\alpha \right) = p \left( \bigwedge_{i \neq j} r_j > r_i | \alpha \right) \; .
\end{equation}

Or, using the integral notation from equation (\ref{eq:EP-def}):

\begin{equation} \label{eq:Dir-EP-int}
\varphi_j = \int_{r_j = \mathrm{max}[r]} \mathrm{Dir}(r; \alpha) \, \mathrm{d}r \; .
\end{equation}

\subsection{Case I: Bivariate Dirichlet} \label{sec:Case_I}

If $k = 2$, the Dirichlet distribution reduces to

\begin{equation} \label{eq:Dir2-pdf}
p(r) = \frac{\Gamma(\alpha_1 + \alpha_2)}{\Gamma(\alpha_1) \Gamma(\alpha_2)} \, r_1^{\alpha_1-1} \, r_2^{\alpha_2-1}
\end{equation}

and therefore becomes a beta distribution

\begin{equation} \label{eq:Beta-pdf}
p(r_1) = \frac{r_1^{\alpha_1-1} \, (1-r_1)^{\alpha_2-1}}{\mathrm{B}(\alpha_1,\alpha_2)}
\end{equation}

with the beta function given by

\begin{equation} \label{eq:beta-fct}
\mathrm{B}(\alpha,\beta) = \frac{\Gamma(\alpha) \, \Gamma(\beta)}{\Gamma(\alpha + \beta)} \; .
\end{equation}

Thus, the exceedance probability for this bivariate case simplifies to

\begin{equation} \label{eq:Dir2-EP-def}
\varphi_1 = p(r_1 > r_2) = p(r_1 > 1 - r_1) = p(r_1 > \sfrac{1}{2}) = \int_{\frac{1}{2}}^1 p(r_1 ) \, \mathrm{d}r_1 \; .
\end{equation}

Using the beta cumulative distribution function, it evaluates to

\begin{equation} \label{eq:Dir2-EP}
\varphi_1 = 1 - \frac{\mathrm{B}(\frac{1}{2};\alpha_1,\alpha_2)}{\mathrm{B}(\alpha_1,\alpha_2)}
\end{equation}

with the incomplete beta function

\begin{equation} \label{eq:inc-beta-fct}
\mathrm{B}(x;\alpha,\beta) = \int_0^x x^{\alpha-1} \, (1-x)^{\beta-1} \, \mathrm{d}x \; .
\end{equation}

As one can see, Dirichlet exceedance probabilities becomes particularly intuitive when $k = 2$, because the statement that $r_1 > r_2$ is equivalent to the statement that $r_1 > \sfrac{1}{2}$.

\pagebreak
\subsection{Case II: Multivariate Dirichlet} \label{sec:Case_II}

If $k > 2$, exceedance probabilities are less intuitive, because in general

\begin{equation} \label{eq:Dir-EP-ineq}
\varphi_j = p(r_j = \mathrm{max}[r]) > p(r_j > \sfrac{1}{2}) \quad \text{for} \quad j = 1, \ldots, k \; ,
\end{equation}

i.e. exceedance probabilities cannot be evaluated using a simple threshold on $r_j$, because $r_j$ might be the maximal element in $r$ without being larger than $\sfrac{1}{2}$. In fact, with $k = 100$, $r_j$ can be $\mathrm{max}[r]$ with just being slightly larger than $\sfrac{1}{100}$. In order to account for this, two approaches can be taken.

\vspace{0.5em}

Using the first method, exceedance probabilities are calculated via sampling from the respective distribution. Dirichlet random numbers can be generated by first drawing $q_1, \ldots, q_k$ from independent gamma distributions with shape parameters $\alpha_1, \ldots, \alpha_k$ and rate parameters $\beta_1 = \ldots = \beta_k$ and then dividing each $q_j$ by the sum over all $q_j$ (Gelman et al., 2013, p.~583). This makes use of the relation

\begin{equation} \label{eq:Gam-Dir-AB}
\begin{split}
& Y_1 \sim \mathrm{Gam}(\alpha_1,\beta), \, \ldots, \, Y_k \sim \mathrm{Gam}(\alpha_k,\beta), \, Y_s = \sum_{j=1}^k Y_j \\
\Rightarrow \quad & X = (X_1, \ldots, X_k) = \left( \frac{Y_1}{Y_s}, \ldots, \frac{Y_k}{Y_s} \right) \sim \mathrm{Dir}(\alpha_1, \ldots, \alpha_k)
\end{split}
\end{equation}

where the probability density function of the gamma distribution is given by

\begin{equation} \label{eq:Gam-pdf}
\mathrm{Gam}(y; a, b) = \frac{{b}^{a}}{\Gamma(a)} \, y^{a-1} \, \exp[-b y] \quad \text{for} \quad y > 0 \; .
\end{equation}

Upon random number generation, exceedance probabilities can be estimated as

\begin{equation} \label{eq:Dir-EP1}
\varphi_j = \frac{1}{S} \sum_{n=1}^{S} \left[ \bigwedge_{i \neq j} r_j^{(n)} > r_i^{(n)} \right]
\end{equation}

where $[ \ldots ]$ refers to Iverson bracket notation, $S$ is the number of samples and $r_j^{(n)}$ corresponds to the $j$-th element from the $n$-th sample of $r$. Unfortunately, sampling is time-consuming and precise estimation of Dirichlet exceedance probabilities requires up to $10^6$ samples. We therefore propose another method relying on numerical integration.

\vspace{0.5em}

Using this second method, exceedance probabilities are again calculated using theorem (\ref{eq:Gam-Dir-AB}). Therefore, consider

\begin{equation} \label{eq:Gam-Dir-A}
q_1 \sim \mathrm{Gam}(\alpha_1,1), \, \ldots, \, q_k \sim \mathrm{Gam}(\alpha_k,1), \, q_s = \sum_{j=1}^k q_j
\end{equation}

and the Dirichlet variate

\begin{equation} \label{eq:Gam-Dir-B}
r = (r_1, \ldots, r_k) = \left( \frac{q_1}{q_s}, \ldots, \frac{q_k}{q_s} \right) \sim \mathrm{Dir}(\alpha_1, \ldots, \alpha_k) \; .
\end{equation}

\pagebreak
Obviously, it holds that

\begin{equation} \label{eq:Gam-Dir-eq}
r_j > r_i \; \Leftrightarrow \; q_j > q_i \quad \text{for} \quad i,j = 1, \ldots, k \quad \text{with} \quad i \neq j \; .
\end{equation}

Therefore, consider the probability that $q_j$ is larger than $q_i$, given $q_j$ is known. This probability is equal to the probability that $q_i$ is smaller than $q_j$, given $q_j$ is known

\begin{equation} \label{eq:Gam-EP0}
p(q_j > q_i|q_j) = p(q_i < q_j|q_j)
\end{equation}

which can be expressed in terms of the gamma cumulative distribution function as

\begin{equation} \label{eq:Gam-EP1}
p(q_i < q_j|q_j) = \int_0^{q_j} \mathrm{Gam}(q_i;\alpha_i,1) \, \mathrm{d}q_i = \frac{\gamma(\alpha_i,q_j)}{\Gamma(\alpha_i)}
\end{equation}

where $\Gamma(\alpha)$ is the gamma function and $\gamma(\alpha,x)$ is the lower incomplete gamma function. Since the gamma variates are independent of each other, these probabilties factorize:

\begin{equation} \label{eq:Gam-EP2}
p(\forall_{i \neq j} \left[ q_j > q_i \right]|q_j) = \prod_{i \neq j} p(q_j > q_i|q_j) = \prod_{i \neq j} \frac{\gamma(\alpha_i,q_j)}{\Gamma(\alpha_i)} \; .
\end{equation}

Although it can be easily calculated using implementations of the gamma function and the lower incomplete gamma function in numerical software packages, this probability is still dependent on $q_j$. In order to obtain the exceedance probability $\varphi_j$, $q_j$ has to be integrated out. From equations (\ref{eq:Dir-EP-def}) and (\ref{eq:Gam-Dir-eq}), it follows that

\begin{equation} \label{eq:Dir-EP2a}
\varphi_j = p(\forall_{i \neq j} \left[ r_j > r_i \right]) = p(\forall_{i \neq j} \left[ q_j > q_i \right]) \; .
\end{equation}

Using the law of marginal probability, we have

\begin{equation} \label{eq:Dir-EP2b}
\varphi_j = \int_0^\infty p(\forall_{i \neq j} \left[ q_j > q_i \right]|q_j) \, p(q_j) \, \mathrm{d}q_j \; .
\end{equation}

With (\ref{eq:Gam-EP2}) and (\ref{eq:Gam-Dir-A}), this becomes

\begin{equation} \label{eq:Dir-EP2c}
\varphi_j = \int_0^\infty \prod_{i \neq j} \left( p(q_j > q_i|q_j) \right) \, \mathrm{Gam}(q_j;\alpha_j,1) \, \mathrm{d}q_j \; .
\end{equation}

And with (\ref{eq:Gam-EP1}) and (\ref{eq:Gam-pdf}), it becomes

\begin{equation} \label{eq:Dir-EP2}
\varphi_j = \int_0^\infty \prod_{i \neq j} \left( \frac{\gamma(\alpha_i,q_j)}{\Gamma(\alpha_i)} \right) \, \frac{q_j^{\alpha_j-1} \exp[-q_j]}{\Gamma(\alpha_j)} \, \mathrm{d}q_j \; .
\end{equation}

In other words, the exceedance probability is an integral from zero to infinity where the first term in the integrand conforms to a product of gamma cumulative distribution functions and the second term is a gamma probability density function.

To our knowledge, the integral in (\ref{eq:Dir-EP2}) cannot be solved analytically. This means that Dirichlet exceedance probabilities for $k > 2$ must be calculated using numerical integration. In the next section, we provide MATLAB code for this task.

\pagebreak
\subsection{MATLAB code} \label{sec:MATLAB_code}

The following code separately handles the cases $k = 2$ and $k > 2$ and is already optimized to avoid overflow and underflow due to very large or very small numbers in the latter case. The \verb|integral| function used in line 27 was introduced in MATLAB R2012a, so this code should work from this version onwards.

\vspace{0.5em}

\small
\begin{verbatim}
01  function exc_p = Dir_exc_prob(alpha)
02  % _
03  % Exceedance Probability for Dirichlet-distributed random variables
04  % FORMAT exc_p = Dir_exc_prob(alpha)
05  %     alpha - a 1 x K vector with Dirichlet concentration parameters
06  %     exc_p - a 1 x K vector with Dirichlet exceedance probabilities
07  
08  % Get dimensionality
09  %-----------------------------------------------------------------------%
10  K = numel(alpha);
11  
12  % Analytical computation, if bivariate Dirichlet
13  %-----------------------------------------------------------------------%
14  if K == 2
15      % using the Beta CDF
16      exc_p(1) = 1 - betainc(1/2,alpha(1),alpha(2));
17      exc_p(2) = 1 - exc_p(1);
18  end;
19  
20  % Numerical integration, if multivariate Dirichlet
21  %-----------------------------------------------------------------------%
22  if K > 2
23      % using Gamma CDFs
24      exc_p = zeros(1,K);
25      for j = 1:K
26          f = @(x) integrand(x,alpha(j),alpha([1:K]~=j));
27          exc_p(j) = integral(f,0,alpha(j)) + integral(f,alpha(j),Inf);
28      end;
29  end;
30  
31  % Integrand function for numerical integration
32  %-----------------------------------------------------------------------%
33  function p = integrand(x,aj,ak)
34  
35  % product of Gamma CDFs
36  p = ones(size(x));
37  for k = 1:numel(ak)
38      p = p .* gammainc(x,ak(k));
39  end;
40  
41  % times a Gamma PDF
42  p = p .* exp((aj-1).*log(x) - x - gammaln(aj));
\end{verbatim}
\normalsize

\pagebreak
\section{Application}

\subsection{Political election forecasting} \label{sec:App_PEF}

In a political election, people can -- if they go to the election and if they do not invalidate their ballot paper -- usually decide between a finite number $k$ of non-overlapping parties. This means that political parties can be understood as event types $E_1, \ldots, E_k$ "occuring" with certain frequencies $r_1, \ldots, r_k$ in the voting population. Since (\ref{eq:Dir-const}) holds true when we restrict ourselves to the voting population filling valid ballot papers, $r$ might be modelled using a Dirichlet distribution given by (\ref{eq:Dir-pdf}), especially when making inferences about $r$ before the election.

Inferences about $r$ before the election are usually based on voting polls. The following table displays voting polls for two elections in the Federal Republic of Germany that happened in 2005 and 2013. The first survey was performed by \textit{Forschungsgruppe Wahlen} on behalf of the public TV service \textit{ZDF} and was published on 09/09/2005 (nine days before the actual election). The second survey was conducted by \textit{Infratest dimap} on behalf of the public TV service \textit{ARD} and was published on 10/01/2013 (ten days before the actual election).

\begin{table}[h]
\centering
\begin{tabular}{lccccccc}
Election & Respondents & CDU & SPD & FDP & Gr\"{u}ne & Linke & Other \\
\hline
\parbox[t]{3.5cm}{Federal election, \\ Germany 2005}
& N = 1,299 & 41 \% & 34 \% & 7 \% &  7 \% & 8 \% & 3 \% \\
\parbox[t]{3.5cm}{State election, \\ Lower Saxony 2013}
& N = 1,001 & 40 \% & 33 \% & 5 \% & 13 \% & 3 \% & 6 \% \\
\hline
\end{tabular}
\end{table}

Voting polls can be considered as data acquisition for a Multinomial-Dirichlet model. Each respondent corresponds to one observation $i = 1,\ldots,n$ and respondents choose between mutually exclusive and collectively exhaustive options $j = 1,\ldots,k$. This is done in order to infer the unknown frequencies $r$ of these options in the whole population. Suppose that option $j$ was selected $n_j$ times in a poll with $\sum_{j=1}^k n_j = n$ and prior knowledge about $r$ is given by $\alpha_0$. Then, the posterior distribution over $r$ is given by $\alpha_{nj} = \alpha_{0j} + n_j$ and informs us about the likelihood of each frequency combination in the population that the respondents were sampled from.

We asssume that the polling results were not corrected for selection biases (which however is usually the case) and that non-voters were excluded from the polling results (such that percentages add up to 100~\%). Furthermore, we apply a uniform prior distribution over $r$ given by $\alpha_0 = [1,\ldots,1]$. Therefore, posterior parameter estimates are given by $\alpha_{nj} = 1 + \sfrac{p_j}{100} \cdot N$ in each poll where $p_j$ refers to the percentage reported for party $j$ in the polling results. Rounded alpha estimates are given in the following table.

\begin{table}[h]
\centering
\begin{tabular}{lcccccc}
Election & CDU & SPD & FDP & Gr\"{u}ne & Linke & Other \\
\hline
2005 & 534 & 443 &  92 &  92 & 105 &  40 \\
2013 & 401 & 331 &  51 & 131 &  31 &  61 \\
\hline
\end{tabular}
\end{table}

These alpha parameters define a Dirichlet distribution over voting frequencies that can be used to make posterior inference by quantifying the probability of posterior statements like $r_{\mathrm{CDU}} > r_{\mathrm{SPD}}$ that are of interest when predicting the outcome of the election. Calculating exceedance probabilities for such a posterior distribution is equivalent to calculating the probability that a certain party will win the election in terms of receiving the maximum number of votes among all the parties.

For the first election (Germany, 2005), we calculate exceedance probabilities for each party. For the second election (Lower Saxony, 2013), we calculate exceedance probabilities for the three blocks "CDU \& FDP" (forming a center-right coalition), "SPD \& Gr\"{u}ne" (forming a center-left coalition) and "Linke \& Other" (the remaining parties). This is done by marginalizing the posterior distribution with respect to these groups based on the agglomeration theorem for the Dirichlet distribution

\begin{equation} \label{eq:Dir-agglom}
(r_1, \ldots, r_k) \sim \mathrm{Dir}(\alpha_1, \ldots, \alpha_k) \Rightarrow \left( \sum_{j \in S_1} r_j, \ldots, \sum_{j \in S_l} r_j \right) \sim \mathrm{Dir} \left( \sum_{j \in S_1} \alpha_j, \ldots, \sum_{j \in S_l} \alpha_j \right)
\end{equation}

where $S_1,\ldots,S_l$ are disjoint subsets of $S = \left\lbrace 1,\ldots,k \right\rbrace$ partitioning the whole set of options, just like it is done by the three political groups mentioned above. Upon marginalization, exceedance probabilities are calculated for the newly obtained distribution. In this way, we make inferences not about which party will win the election, but about which part of the political spectrum will most likely be able to form a coalition. In the following table, we report these exceedance probabilities.

\begin{table}[h]
\centering
\begin{tabular}{lcccccc}
Election & CDU & SPD & FDP & Gr\"{u}ne & Linke & Other \\
\hline
2005 & 99.82 \% & 0.18 \% & 0 \% & 0 \% & 0 \% & 0 \% \\
\hline
\\
Election & \multicolumn{2}{c}{CDU \& FDP} & \multicolumn{2}{c}{SPD \& Gr\"{u}ne} & \multicolumn{2}{c}{Linke \& Other} \\
\hline
2013 & \multicolumn{2}{c}{37.04 \%} &  \multicolumn{2}{c}{62.96 \%} & \multicolumn{2}{c}{0.00 \%} \\
\hline
\end{tabular}
\end{table}

This means that, according to the voting polls listed above, the CDU was most likely to win the federal election in 2005 and the center-left wing parties were more likely though far from certain to win the state election in 2013. We list actual election results in the following table.

\begin{table}[h]
\centering
\begin{tabular}{lcccccc}
Election & CDU & SPD & FDP & Gr\"{u}ne & Linke & Other \\
\hline
2005 & 35.2 \% & 34.2 \% &  9.8 \% &  8.1 \% &  8.7 \% &  4.0 \% \\
2013 & 36.0 \% & 32.6 \% &  9.9 \% & 13.7 \% &  3.1 \% &  4.7 \% \\
\hline
\end{tabular}
\end{table}

Thus, with regard to the final results, both predictions were correct, but only with some reservations: In 2005, the CDU won the election by a much closer margin than expected which lead to heavy criticism of forecasting institutes; and in 2013, the race between the center-right and the center-left was too close to call until late into the election night which is also reflected in the lower maximal exceedance probability in this comparison.

\pagebreak
\subsection{Neuroimaging model selection} \label{sec:App_NMS}

Consider Bayesian inference on data $y$ using model $m$ with parameters $\theta$. Then, $p(y|m)$ is the probability of the data $y$, given only the model $m$, regardless of particular values of the parameters $\theta$ which are integrated out of the likelihood function $p(y|\theta,m)$. This probability is called "model evidence" or "marginal likelihood" and can act as a model quality criterion in Bayesian inference. In neuroimaging, a hierarchical model has been proposed (Stephan et al., 2009) that allows to make inferences on model frequencies and optimal model structure in a population, given log model evidences $\log \; p(y|m)$ for a number of subjects from this population and a number of models. This is called group-level or random-effects Bayesian model selection (RFX BMS).

We consider model evidences $p(y_i|m_j)$ for the data from subjects $i = 1,\ldots,N$ analyzed using models $j = 1,\ldots,M$. Then, the hierarchical population proportion model underlying RFX BMS is given by the following probability densities:

\begin{equation} \label{eq:BMS-RFX}
\begin{split}
p(y|m) &= \prod_{i=1}^N p(y_i|m_i) = \prod_{i=1}^N \prod_{j=1}^M p(y_i|e_j) ^ {m_{ij}} \\
p(m|r) &= \prod_{i=1}^N \mathrm{Mult}(m_i; 1, r) = \prod_{i=1}^N \prod_{j=1}^M r_j ^ {m_{ij}} \\
p(r|\alpha) &= \mathrm{Dir}(r; \alpha) = \frac{\Gamma(\sum_{j=1}^M \alpha_j)}{\prod_{j=1}^M \Gamma(\alpha_j)} \, \prod_{j=1}^M r_j^{\alpha_j-1}
\end{split}
\end{equation}

where $y = \left\lbrace y_1, \ldots, y_N \right\rbrace$ represents measured data, $m$ is an $N \times M$ indicator matrix representing the multinomial variable "model", $r$ is a $1 \times M$ vector of unknown model frequencies in the population and $\alpha$ is a $1 \times M$ vector of concentration parameters.

Accounting for different subjects being best explained by different models, a Variational Bayesian (VB) algorithm has been developed (Stephan et al., 2009) to infer a posterior distribution over model frequencies $p(r|y)$ from prior concentration parameters $\alpha_0$:

\begin{equation} \label{eq:BMS-RFX-VB}
\begin{split}
& \alpha = \alpha_0 = [1,\ldots,1]\\
& \text{until convergence} \\
& \quad \quad u_{ij} = \exp \left[ \log \; p(y_i|e_j) + \psi (\alpha_j) - \psi\left(\sum_{j=1}^M \alpha_j\right) \right] \\
& \quad \quad \beta_j = \sum_{i=1}^N \frac{u_{ij}}{u_i}, \; u_i = \sum_{j=1}^M u_{ij} \\
& \quad \quad \alpha = \alpha_0 + \beta \\
& \text{end} \\
& p(r|y) = \mathrm{Dir}(r; \alpha)
\end{split}
\end{equation}

Upon model estimation, exceedance probabilities can be calculated for the Variational posterior Dirichlet distribution in order to make quantitative statements about the "winning model", i.e. the model that best explains a given set of data.

\pagebreak
The model in equation (\ref{eq:BMS-RFX}) can be seen as a simple extension of a Multinomial-Dirichlet model (see Section \ref{sec:App_PEF}) where the occurences of event types are not observed directly, but through model evidences $p(y|m)$.

Originally, RFX BMS was introduced for dynamic causal models (DCMs) (Stephan et al., 2009; Penny et al., 2010; Rigoux et al., 2014) where the log model evidence is approximated using the variational free energy (Friston et al., 2006). However, the approach can also operate on general linear models (GLMs) (Rosa et al., 2010) which requires log model evidences to be calculated voxel-wise (Penny et al., 2007), i.e. separately for each measurement location in the brain.

Here, we apply RFX BMS to a study on orientation pop-out processing (Bogler et al., 2013) where brain activity data was acquired using functional magnetic resonance imaging (fMRI) and analyzed using voxel-wise univariate general linear models (GLMs). We constructed two model spaces, one having three models and the other having nine models. For each model in each of the 22 subjects, we calculated a cross-validated log model evidence (cvLME). For both model spaces, we estimated the RFX BMS model resulting in posterior densities $p(r|\alpha)$. From these posterior densities, we calculated exceedance probabilities $\varphi$ in all 53,268 in-mask voxels.

We compared EP calculation using the numerical integration approach given in equation (\ref{eq:Dir-EP2}) and using the provided code (see Section \ref{sec:MATLAB_code}) against the sampling approach given in equation (\ref{eq:Dir-EP1}) which is currently implemented in the software package \textit{Statistical Parametric Mapping} (SPM) as the function \verb|spm_dirichlet_exceedance.m|. Resulting EP calculation times are given in the following table.

\begin{table}[h]
\centering
\begin{tabular}{lccr}
Model Space & Integration & Sampling & Ratio \\
\hline
3 models & 03:08 min & 33:58 min & 10.84 \\
9 models & 13:44 min & 97:58 min &  7.13 \\
\hline
\end{tabular}
\end{table}

This shows that numerical integration outperforms random sampling by a factor of 7 to 11. We expect that sampling time increases linearly with number of models (more random numbers need to be generated) whereas integration time will grow faster than linear (the integrand gets more and more complex). Therefore, with more models, the advantage of the integration approach might disappear or even reverse. However, with a lot of models in RFX BMS, it is also not reasonable anymore to compute model EPs, but to group models into model families using (\ref{eq:Dir-agglom}) and then calculate family EPs.

Also note that for the sampling approach, we only used $S = 10^5$ samples which is below the recommended number of samples $S = 10^6$ (the default value in the SPM function). Whereas the precision of the sampling approach critically depends on the number of samples, the accuracy of numerical integration is very high, because it only depends on the MATLAB implementation. Here, we used MATLAB R2013b on a 64-bit Windows 7 PC with 16 GB RAM and eight CPU kernels working at 3.40 GHz.

\pagebreak
\section{Conclusion}

We have derived an efficient method to calculate exceedance probabilities for the Dirichlet distribution based on numerical integration over a product of gamma cumulative distribution functions multiplied with a gamma density. Moreover, we have provided MATLAB code for implementing this method and presented two applications.

Using the example of political election forecasting, we have investigated the properties of Dirichlet EPs in their application to survey data. Turning to the example of neuroimaging model selection, we have shown that EP calculation using numerical integration clearly outperforms a sampling approach based on randum number generation which is currently used for model inference in neuroimaging, especially fMRI data analysis.

\section{References}

\renewcommand{\section}[2]{}

\end{document}